\documentclass[12pt]{article}

\usepackage{geometry}
\title{Density-based modeling and identification of biochemical networks in cell populations}
\author{J.~Hasenauer$^{1,*}$, S.~Waldherr$^1$, M.~Doszczak$^2$, \\
  P.~Scheurich$^2$, and F. Allg{\"o}wer$^1$ \\ \\
  $^1$Institute of Systems Theory and Automatic Control\\
  University of Stuttgart, Germany\\
  www.ist.uni-stuttgart.de \\ \\
  $^2$Institute of Cell Biology and Immunology, \\
  University of Stuttgart, Germany\\
  www.uni-stuttgart.de/izi\\ \\
  $^*$Corresponding author (hasenauer@ist.uni-stuttgart.de)
}
\date{February 24, 2010}

\usepackage{natbib}	
\usepackage{graphicx}
\usepackage{amsmath,amssymb,amsbsy}
\usepackage{xcolor}

\newtheorem{problem}{Problem}
\newtheorem{remark}{Remark}

\begin{document}

\maketitle

\section*{Abstract}
In many biological processes heterogeneity within cell populations is an important issue. In this work we consider populations where the behavior of every single cell can be described by a system of ordinary differential equations. Heterogeneity among individual cells is accounted for by differences in parameter values and initial conditions. Hereby, parameter values and initial conditions are subject to a distribution function which is part of the model specification. Based on the single cell model and the considered parameter distribution, a partial differential equation model describing the distribution of cells in the state and in the output space is derived.

For the estimation of the parameter distribution within the model, we consider experimental data as obtained from flow cytometric analysis. From these noise-corrupted data a density-based statistical data model is derived. Using this data model the parameter distribution within the cell population is computed using convex optimization techniques.

To evaluate the proposed method, a model for the caspase activation cascade is considered. It is shown that for known noise properties the unknown parameter distributions in this model are well estimated by the proposed method.
\\[3ex]
\textbf{Keywords:} parameter estimation, cell population, kernel-density estimation, flow cytometry, convex optimization

\section{Introduction}
\label{sec:intro}

Most of the modeling performed in the area of systems biology aims at achieving a quantitative description of intracellular pathways. Hence, most available models describe a ''typical cell'' on the basis of experimental data. Unfortunately, experimental data are in general obtained using cell population experiments, e.g.\ western blotting. If the considered population is highly heterogeneous, meaning that there is a large cell-cell variability, fitting a single cell model to cell population data can lead to biologically meaningless results. To understand the dynamical behavior of heterogeneous cell populations it is crucial to develop integrated cell population models.

Modeling on the population scale has already been addressed by \cite{Mantzaris2007} and \cite{MunskyTri2009}. These authors demonstrated that populations can show a bimodal response if stochasticity in biochemical reactions is considered. But besides stochasticity in biochemical reactions there are other reasons which can also lead to heterogeneity in populations. Examples are unequal partitioning of cellular material at cell division \citep{Mantzaris2007}, genetic and epigenetic differences \citep{Avery2006}.

For the purpose of this paper, we describe heterogeneity in populations by differences in parameter values of the model describing the single cell dynamics. The network structure is assumed to be identical in all cells, as this usually represents the physical interactions among molecules, which should be independent of the cell's state. This parametric approach is well suited for genetic and epigenetic differences. The distribution of parameter values within the cell population of interest is described by a multivariate probability density function, which is part of the model specification.

In the following the problem of estimating the parameter distribution function is studied. Therefore, we consider high-throughput experimental methods such as flow cytometry, which can be used to measure concentration distributions within cell populations by suitable fluorescent labeled antibodies. Classical flow cytometry devices can measure several thousand cells per second.

To estimate the parameter distributions, in a first step, an appropriate population model has to be found. In the literature mathematical models of cell populations are either described as cell ensembles \citep{hasenauer09b,MunskyTri2009}, or as a non-linear partial differential equation (PDE) for the distribution of the state variables \citep{Mantzaris2007,Luzyanina2009,Tsuchiya1966}. In case of ensemble models, a differential equation is assigned to each cell, making an in depth theoretical analysis difficult. PDE models, which describe the time evolution of the distributions of the state variables based on the single cell models, are easy to handle from a theoretical point of view but hard to simulate for a large state dimension of the single cell model. Therefore, only low dimensional PDE models of populations have been studied in literature so far \citep{Mantzaris2007,Luzyanina2007,Luzyanina2009}.

In this paper a PDE model for the state distribution within a heterogeneous cell population is derived. Given the solution of this PDE the probability density of measuring a certain output can be determined. As for the estimation only the measured outputs are required, a numerical method for computing the output distribution is outlined. This methods employs a particle-based approach \citep{Rawlings2006} and classical density estimation \citep{Silverman}. 

Based on these efficient computation scheme for the population response an estimation method is developed. A statistical model of the measured output distribution is derived from the single cell measurement obtained at every measurement instance. Therefore, again kernel density estimators are used as they have better asymptotic properties than commonly used naive estimators \citep{Luzyanina2009}. Given a model and the output distribution estimated from the measurement, a $l_2$-norm minimization is performed over the set of possible parameter distributions. By employing the model properties and a parameterization of the parameter distribution this optimization problem is convex and can be solved efficiently.

The paper is structured as follows. In Section 2, the problem of estimating the parameter distribution is introduced. In Section 3, we present the statistical model for the measured data and the simulation model for state and output distribution. Section 4 gives a short overview of the used identification procedure before in Section 5 the proposed method is applied to a caspase activation model with artificial data.

\textit{Notation:} Consider the $m$-dimensional hypersurface $\mathcal{S} \subset \mathbb{R}^n$. The integral $I$ of a function $f(x)$, with $f:\mathbb{R}^n \rightarrow \mathbb{R}$, over $x \in \mathcal{S}$ is written as
\begin{equation}
\begin{aligned}
I = \int_{\mathcal{S}} f(x) dS.
\end{aligned}
\end{equation}
Furthermore, the $i$.th unit vector is denoted by $e_i$.

\section{Problem statement}
\label{sec:prob}

For the purpose of this work, a model of a biochemical reaction network in a population of $M$ cells is given by the collection of differential equations
\begin{equation}
\begin{aligned}
\dot{x}^{(i)} &= f(x^{(i)},p^{(i)}), \quad x^{(i)}(0) = x_0^{(i)},\\
y^{(i)} &= h(x^{(i)},p^{(i)}), \quad i \in \{1,\ldots,M\} 
\end{aligned}
\label{eq:cell pop}
\end{equation}
with state variables $x^{(i)}(t) \in \mathbb{R}^n$, measured variables $y^{(i)}(t) \in \mathbb{R}^m$, and parameters $p^{(i)} \in \mathbb{R}^q$. The index $i$ specifies the individual cells within the population. The parameters $p^{(i)}$ can be kinetic constants, e.g. reaction rates or binding affinities. The cell-cell interaction of the considered pathway is assumed to be negligible, as it is the case in many \textit{in vitro} lab experiments. 

In the following heterogeneity within the cell population is introduced, modeled by differential parameter values and initial conditions among individual cells. The distribution of parameters $p^{(i)}$ and initial conditions $x_0^{(i)}$ is given by a probability density function $\Phi: \mathbb{R}^{n+q} \rightarrow \mathbb{R}_+$ with $\int_{\mathbb{R}^{n+q}} \Phi(x_0,p) dx_0 dp = 1$. For ease of notation, we write $\xi_0 = (x_0 ^T, p ^T) ^T$. The probability density function $\Phi$ is part of the model specification and the parameters and initial conditions of cell $i$ are subject to the probability distribution
\begin{equation}
\begin{aligned}
\mathrm{\mathop{Pr}}(\xi_{0,1}^{(i)} \leq \xi_1,\cdots,\xi_{0,n+q}^{(i)} \leq \xi_{n+q}) =  \int_{-\infty}^{\xi_1} \hspace{-3mm} \cdots \int_{-\infty}^{\xi_{n+q}} \Phi(\tilde{\xi}) d\tilde{\xi}_1 \cdots d\tilde{\xi}_{n+q}.
\end{aligned}
\label{eq: parameter distribution}
\end{equation}

As outlined in Section \ref{sec:intro}, for the study of cell populations high-throughput cell population measurements are available. Using these experimental techniques protein concentrations within thousands of cells can be measured at every measurement instance, $t_k$, $k = 1,\ldots,N$. This yields the measurement data
\begin{equation}
\mathcal{D}_k = \left\{\left(t_k,\psi^{(i)}(t_k)\right)\right\}_{i \in \mathcal{I}_k}, \quad k = 1,\ldots,N
\label{eq: data}
\end{equation}
where $\psi^{(i)}$ is the measured output of the cell $i$ and $\mathcal{I}_k$ is the index set of the cells measured at time $t_k$. Note that the cells cannot be tracked over time, and are removed from the population in order to obtain the measurements. Thus, no single-cell time series data are available. On the other hand, the samples are independent and equally distributed and $\mathrm{card}(\mathcal{I}_k)$ is assumed to be large, such that an approximation of the output distribution is possible. 

Like most measurement devices, also high-throughput fluorescence measurements are subject to noise.
For the rest of the paper, noise consisting of a relative and an absolute part is considered,
\begin{equation}
\psi^{(i)}(t_k) = \mathrm{diag}(\eta^1) y^{(i)}(t_k) + \eta^2,
\label{eq: measurement}
\end{equation}
in which $\psi^{(i)}$ is the measured output and $\eta^j \in \mathbb{R}^m$ is a vector of $\mathrm{log}$-normally distributed random variables with probability density functions
\begin{equation}
\Theta^i_j(\eta^i_j) = \dfrac{1}{\sqrt{2 \pi} \sigma^i_j \eta^i_j} \exp \left\{ -\dfrac{1}{2}\left(\dfrac{\log \eta^i_j - \mu^i_j}{\sigma^{i}_j}\right)^2\right\}, 
\quad i = 1,2, \quad j=1,\dotsc,m,
\label{eq: pdf log-normal}
\end{equation}
yielding the joint probability density
\begin{equation}
\Theta^i(\eta^i) = \prod_{j=1}^m \Theta^i_j(\eta^i_j).
\label{eq: joint pdf log-normal}
\end{equation}
$\mathrm{Log}$-normally distributed random variables are chosen here, since they are a good model for the commonly seen noise distributions of the considered measurement device and conserve the positivity of all variables. For notational simplicity the measurement errors of the different concentrations are assumed to be uncorrelated. This constraint can be removed easily.

Given this setup the problem we are concerned with is:
\begin{problem}
Given the measurement data $\mathcal{D}_k$, $k = 1,\ldots,N$, the cell population model \eqref{eq:cell pop}, and the noise model \eqref{eq: joint pdf log-normal}, determine the parameter distribution $\Phi(\xi)$.
\end{problem}

Unfortunately, estimation of $\Phi(\xi)$ using a cell population model with a finite number of cells and discrete sampled data is fairly difficult as no single cell trajectories are available. A far more natural approach would be to use a density description, as the available measurement data can be interpreted as samples drawn from the probability density function of the output. This interpretation is also quite appealing from a point of modeling as the number of cells considered in a standard lab experiment is of the order of $10^9$ and hence nevertheless too large to be simulated on an individual basis. In the next chapter a PDE model for the probability density of the output and a density model for the measurement data is derived.

\section{Density-based modeling of heterogeneous cell populations}
\label{sec:modeling}
As outlined in the previous section, a continuous statistical model for the measurement data, as well as for the evolution of the state and output density would be preferable. These two aspects are addressed in the following.

\subsection{Density model of measurement data}
\label{subsec:data modeling}
The data collected by the considered measurement devices $\mathcal{D}_k$ are samples drawn from the distribution of the measured output, as mentioned in Section \ref{sec:prob}. Let $\Psi(\psi,t_k)$ be the distribution of the measured outputs $\psi^{(i)}(t_k)$ at time $t_k$. As $\Psi(\psi,t_k)$ is considered to be a probability density, classical density estimation methods can be employed for estimating $\Psi(\psi,t_k)$ from the given samples $\mathcal{D}_k$.

\begin{figure}[t!]
\begin{center}
\includegraphics{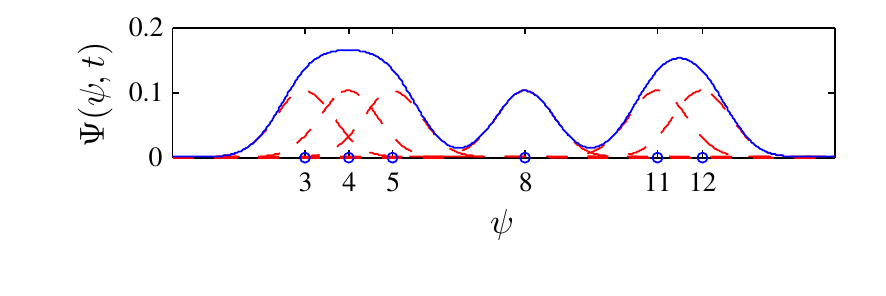}
\end{center}
\caption{Gaussian kernel density estimate ({\color{blue}---}) of $\Psi(\psi,t)$ for the measured outputs ({\color{blue}o}) and the associated Gaussian kernels ({\color{red}-- --}).}
\label{fig:kde}
\end{figure}

In this work, the problem of determining $\Psi(\psi,t_k)$ from $\mathcal{D}_k$ is approached using kernel density estimators. Kernel density estimators are non-parametric approaches to estimate probability distributions from sampled data \citep{Silverman}. They are widely used and can be thought of as placing probability ''bumps'' at each observation, as depicted in Figure \ref{fig:kde}. These ''bumps'' are the kernel function $K$, with $\int_{\mathbb{R}^m} K(\psi) d\psi = 1$. Note that here only the equations for the one dimensional case are given. The extension towards higher dimensions is straightforward and can be found in \cite{Silverman}. In this work, a Gaussian kernel given by
\begin{equation}
\begin{aligned}
K\left(\psi-\psi^{(i)},{h}\right) = \dfrac{1}{\sqrt{2 \pi} h} \exp \left\{ -\dfrac{1}{2} \left(\dfrac{\psi-\psi^{(i)}}{h} \right)^2\right\},
\end{aligned}
\label{eq:kernel function}
\end{equation}
with standard deviation $h$ is used. In this context, $h$ is also called smoothing parameter in the literature \citep{Silverman}.

Given the kernel $K$ an estimator of the probability density for a given set of samples $\mathcal{D}_k$ is
\begin{equation}
\begin{aligned}
\Psi(\psi,t_k) = \frac{1}{M_k}\sum_{i\in\mathcal{I}_k} K\left(\psi-\psi^{(i)}(t_k),h\right),
\end{aligned}
\label{eq:estimator for measured output density}
\end{equation}
where $M_k$ is the cardinality of $\mathcal{I}_k$. The selection of the smoothing parameter $h$ is crucial and depends strongly on $M_k$. In this work $h$ is chosen according to the least-squares cross-validation method \citep{Stone1984}. As $M_k$ is considered to be of order $10^4$, it can be assumed that the the estimated output distribution ins close to the actual output distribution. 

\subsection{PDE model of density evolution}
As outlined previously, a continuous model for the output density is desirable for the purpose of parameter identification. Therefore, a PDE model for the cell population is derived in the next step.

At first the single cell model is transformed in an extended state space model
\begin{equation}
\begin{aligned}
\dot{\xi}^{(i)}&=
\left(\begin{array}{cc}
f(\xi^{(1,i)},\xi^{(2,i)}) \\ 0
\end{array}\right),
\quad
\xi^{(i)}(0) =
\left(\begin{array}{cc}
x_0^{(i)} \\ p^{(i)}
\end{array}\right)
\\
y^{(i)} &= h(\xi^{(1,i)},\xi^{(2,i)})
\end{aligned}
\end{equation}
in which the parameters are appended to the state vector, $\xi^{(i)} = [\xi^{(1,i)},\xi^{(2,i)}]^T \in \mathbb{R}^{n+q}$ with $\xi^{(1,i)} = x^{(i)}$ and $\xi^{(2,i)} = p^{(i)}$. This system can also be written as
\begin{equation}
\begin{aligned}
\dot{\xi}^{(i)} &= F(\xi^{(i)}), \quad \xi^{(i)}(0) = \xi_0^{(i)}\\
y^{(i)} &= H(\xi^{(i)}),
\end{aligned}
\label{eq: single cell reformulated}
\end{equation}
to which we refer as the extended state space representation.

\begin{figure}[t!]
\begin{center}
\includegraphics[width=7cm]{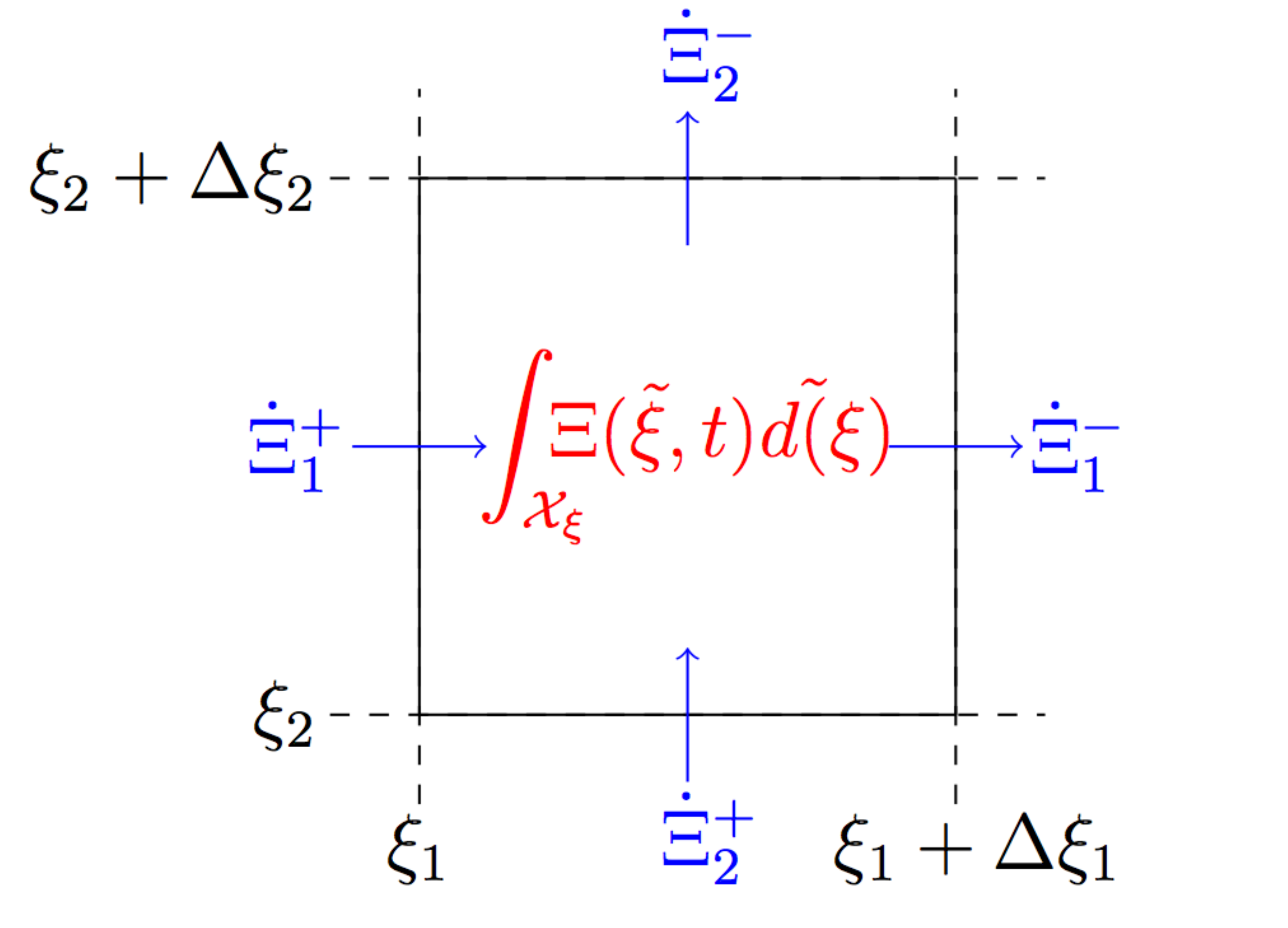}
\end{center}
\caption{Infinitesimal volume element $\mathcal{X}$ of the extended state space, with fluxes across the boundaries.}
\label{fig:FEM}
\end{figure}

Based on \eqref{eq: single cell reformulated}, the PDE model for the population is derived. The state variable of this PDE is the state distribution function $\Xi:\mathbb{R}^{n+q}\times\mathbb{R}\rightarrow\mathbb{R}_+:(\xi,t)\mapsto\Xi(\xi,t)$, which is defined on the extended state space. Based on the distribution function $\Xi$, the probability of picking at random a cell from the population with states $\xi^{(i)}(t) \in \mathcal{X}$ at time $t$ is given by
\begin{equation}
\mathrm{\mathop{Pr}}(\xi^{(i)}(t) \in \mathcal{X}) = \int_\mathcal{X} \Xi(\tilde{\xi},t) d\tilde{\xi}.
\end{equation}
To determine the PDE for $\Xi$, an infinitesimal volume $\mathcal{X}_\xi = \mathcal{X}_{\xi,1}\times\ldots\times\mathcal{X}_{\xi,n+q}$ of the extended state space is considered, with $\mathcal{X}_{\xi,i} = [\xi_i,\xi_i + \Delta \xi_i]$. For the 2-dimensional case this is depicted in Figure~\ref{fig:FEM}.

For this infinitesimal volume the flux and storage balance is,
\begin{equation}
\begin{aligned}
\int_{\mathcal{X}_\xi} \hspace{-2mm} \Xi(\tilde{\xi},t+\Delta t) d\tilde{\xi} -  \int_{\mathcal{X}_\xi} \hspace{-2mm} \Xi(\tilde{\xi},t) d\tilde{\xi} = 
\sum_{i=1}^N \int_{t}^{t+\Delta t} \left( \dot{\Xi}_i^+(\xi,\tau) - \dot{\Xi}_i^-(\xi,\tau) \right) d\tau.
\end{aligned}
\label{eq:flux-balance}
\end{equation}
The left hand side of the equation represents the storage term and the right hand side the fluxes across the boundaries. The fluxes $\dot{\Xi}_i^+$ and $\dot{\Xi}_i^-$ are given by the surface integral of the product of boundary distribution and entering velocity, determined by the single cell dynamics,
\begin{equation}
\begin{aligned}
\dot{\Xi}_i^+(\xi,t) &= \int_{\mathcal{S}_{\dot{\Xi}}(\xi,i)} \hspace{-7mm} F_i(\tilde{\xi}) \Xi(\tilde{\xi},t) dS \\
\dot{\Xi}_i^-(\xi,t) &= \int_{\mathcal{S}_{\dot{\Xi}}(\xi+e_i\Delta \xi_i,i)} \hspace{-16mm} F_i(\tilde{\xi}) \Xi(\tilde{\xi},t) dS,
\end{aligned}
\label{eq:flux}
\end{equation}
in which $\mathcal{S}_{\dot{\Xi}}(\xi,i) = \lbrace\tilde{\xi}|\tilde{\xi}_i = \xi_i \wedge \tilde{\xi}_j \in \mathcal{X}_{j} \forall j \neq i\rbrace$.

Next, \eqref{eq:flux-balance} and \eqref{eq:flux} are used to derive the PDE for the time evolution of $\Xi(\xi,t)$. Therefore, at first the storage term is expanded using its Taylor series, yielding
\begin{equation}
\begin{aligned}
\int_{\mathcal{X}_\xi} \Xi(\tilde{\xi},t+\Delta t) d\tilde{\xi} -  \int_{\mathcal{X}_\xi} \Xi(\tilde{\xi},t) d\tilde{\xi} = 
\left(\Xi(\xi,t+\Delta t) - \Xi(\xi,t)\right)\prod_{j=1}^N \Delta \xi_j + \mathcal{O}(\Delta \xi^{N+1}).
\label{eq:approx storage}
\end{aligned}
\end{equation}
Here it is assumed that $\mathcal{O}(\Delta \xi_j)=\mathcal{O}(\Delta \xi)$ $\forall j \in \{1,\ldots,n+q\}$. In a second step the flux difference $\Delta \dot{\Xi}_i(\xi,\tau) = \dot{\Xi}_i^+(\xi,\tau) - \dot{\Xi}_i^-(\xi,\tau)$ is rewritten,
\begin{equation}
\begin{aligned}
\Delta \dot{\Xi}_i(\xi,t) &= - \int_{\mathcal{S}_{\dot{\Xi}}(\xi,i)} \hspace{-0mm} \Bigl(\left.\frac{\partial(F_i\Xi)}{\partial \xi_i}\right|_{(\xi,t)} \hspace{-4mm} \Delta \xi_i+ \mathcal{O}(\Delta \xi_i^2) \Bigr) dS \\
&= - \left.\frac{\partial(F_i\Xi)}{\partial \xi_i}\right|_{(\xi,t)} \prod_{j=1}^N \Delta \xi_j + \mathcal{O}(\Delta \xi^{N+1}).
\label{eq:approx flux}
\end{aligned}
\end{equation}
The first line follows from the definition of $\Delta \dot{\Xi}_i(\xi_i,t)$ and the Taylor series expansion of $F_i(\xi + e_i \Delta \xi_i) \Xi(\xi + e_i \Delta \xi_i,t)$. To obtain the second line the integration is carried out. The final reformulation is the expansion of the time integral in \eqref{eq:flux-balance}, resulting in
\begin{equation}
\begin{aligned}
\int_{t}^{t+\Delta t} \left( \dot{\Xi}_i^+(\xi,\tau) - \dot{\Xi}_i^-(\xi,\tau) \right) d\tau
= - \Delta t\left(\prod_{j=1}^N \Delta \xi_j\right) \left.\frac{\partial(F_i\Xi)}{\partial \xi_i}\right|_{(\xi,t)} \hspace{-3mm}+ \mathcal{O}(\Delta \xi^N)\mathcal{O}(\Delta t^2).
\label{eq:approx time integral}
\end{aligned}
\end{equation}
Substituting \eqref{eq:approx storage} and \eqref{eq:approx time integral} in the flux balance \eqref{eq:flux-balance} and dividing by $\Delta t \prod_{j=1}^N \Delta \xi_j$ then yields,
\begin{equation}
\begin{aligned}
\hspace{-1mm} \frac{\Xi(\xi,t+\Delta t) - \Xi(\xi,t) + \mathcal{O}(\Delta \xi)}{\Delta t} = -\sum_{i=1}^N \left.\frac{\partial(F_i\Xi)}{\partial \xi_i}\right|_{\xi} \hspace{-2mm} + \mathcal{O}(\Delta t).
\label{eq:flux-balance2}
\end{aligned}
\end{equation}
Given this the PDE governing the evolution of $\Xi(\xi,t)$ is obtained by taking the limits $\Delta \xi_i \rightarrow 0$ and $\Delta t \rightarrow 0$, leading to
\begin{equation}
\begin{aligned}
\frac{\partial \Xi}{\partial t}(\xi,t) &= -\sum_{i=1}^N \frac{\partial(F_i\Xi)}{\partial \xi_i}(\xi,t),
\label{eq:Theta dynamics}
\end{aligned}
\end{equation}
for sufficiently smooth $\Xi(\xi,t)$. This final equation is somehow what we expected, a transport equation with position dependent transport direction and velocity, according to the single cell dynamics. The initial condition of \eqref{eq:PDE} is the initial distribution on the extended state space,
\begin{equation}
\begin{aligned}
\Xi(\xi,0) = \Phi(\xi), \quad \forall \xi \in \mathbb{R}^{n+q}_+.
\label{eq:IC PDE}
\end{aligned}
\end{equation}

From the state distribution $\Xi$, the output distribution $\Upsilon$ is computed as the integral of the state distribution along $H(\xi) = y$,
\begin{equation}
\begin{aligned}
\Upsilon(y,t) = \int_{\mathcal{S}_{\Upsilon}(y)} \hspace{-4mm}\Xi(\xi,t) dS,
\label{eq: output density}
\end{aligned}
\end{equation}
where $\mathcal{S}_{\Upsilon}(y) = \lbrace \xi | H(\xi) = y\rbrace$.

The resulting partial differential equation system is
\begin{equation}
\begin{aligned}
\frac{\partial \Xi}{\partial t}(\xi,t) &= -\sum_{i=1}^N \frac{\partial(F_i\Xi)}{\partial \xi_i}(\xi,t), \quad 
\Xi(\xi,0) = \Phi(\xi)\\
\Upsilon(y,t) &= \int_{\mathcal{S}_{\Upsilon}(y)} \hspace{-4mm}\Xi(\xi,t) dS,
\label{eq:PDE}
\end{aligned}
\end{equation}
where $\Xi: \mathbb{R}^{n+q}\times\mathbb{R} \rightarrow \mathbb{R}_+$ and $\Upsilon: \mathbb{R}^{m}\times\mathbb{R} \rightarrow \mathbb{R}_+$. This PDE is of first order, quasilinear and known as Liouville's equation. The solution always exists for sufficiently smooth $F(\cdot)$ \citep{EvansPDE}.

As the measurements are noise corrupted, the distribution of measured outputs $\Psi(\psi,t)$ is different from the actual output distribution $\Upsilon(y,t)$. It is defined by
\begin{equation}
\begin{aligned}
\hspace{-1mm} \Psi(\psi,t) \hspace{-1mm} &= \hspace{-2mm} \int_{\mathcal{S}_{\Psi}(\psi)} \hspace{-4mm} \Upsilon(y,t) \Theta^1(\eta^1)\Theta^2(\eta^2) dS,
\label{eq:expected measured output density}
\end{aligned}
\end{equation}
where $\mathcal{S}_{\Psi}(\psi) = \lbrace [y^T,(\eta^1)^T,(\eta^2)^T]^T  | \mathrm{diag}(\eta^1) y + \eta^2 = \psi \rbrace$.

\subsection{Numerical solution of PDE}
In order to study the time evolution of the output distribution $\Upsilon(y,t)$ and the measured output distribution $\Psi(\psi,t)$ equation \eqref{eq:PDE} has to be solved for given $\Phi$. As $\Xi(\xi,t)$ is defined on the $(n+q)$-dimensional space, standard grid based solvers are not able to solve \eqref{eq:PDE} for $n+q > 3$. Theoretically, the methods of characteristics can be used \citep{EvansPDE} but for the high dimensional system we are going to study, also this method is difficult to apply. Instead, a stochastic method is used, which is known from particle filtering \citep{Rawlings2006}.

This stochastic integration method is based on a particle description of the model, which is in our case equivalent to the cell ensemble model \eqref{eq:cell pop}. To compute $\Psi(\psi,t_k)$, at first a set of samples $\{(x_0^{(i)},p^{(i)})\}_{i = 1,\ldots,S}$, is drawn from $\Phi(\xi)$, where $S$ is the number of samples. For this set of samples the single cell model \eqref{eq:estimator for measured output density} is simulated, resulting in a set of simulated outputs $\{(y^{(i)}(t))\}_{i = 1,\ldots,S}$. $y^{(i)}(t)$ is then corrupted by noise according to \eqref{eq: measurement} resulting in $\{(\psi^{(i)}(t))\}_{i = 1,\ldots,S}$. Given this a numerical approximation of $\Psi(\psi,t)$ can be determined using the kernel density estimator described in Section \ref{subsec:data modeling}. This numerical stochastic approximation the output of \eqref{eq:PDE} can be shown to converge as $S \rightarrow \infty$. Hence, the measured output distribution $\Psi(\psi,t_k)$ can be axproximated also for high dimensional nonlinear systems.

\section{Estimation of parameter distributions}
\label{sec:estimation}

As mentioned in Section \ref{sec:prob} the problem studied in this work is the estimation of the parameter distribution $\Phi$ from the data $\mathcal{D}_k$. This problem is approached in the following by minimizing the $l_2$-norm of the model-data mismatch, 
\begin{equation}
\begin{aligned}
J\left(\hat{\Phi}\right) = \sum_{k=1}^N \left|\left| \Psi(\psi,t_k) - \hat{\Psi}(\psi,t_k,\hat{\Phi})\right|\right|_2^2.
\end{aligned}
\label{eq:l2 norm}
\end{equation}
in which $\hat{\Psi}(\psi,t,\hat{\Phi})$ is the distribution of the measured output $\psi$ obtained by simulation with the parameter distribution $\hat{\Phi}(\xi)$. According to the cost $J$, the optimal parameter distribution $\hat{\Phi}^*(\xi)$ is than given by \begin{equation}
\begin{array}{l}
\hat{\Phi}^* = \mathrm{arg} \min_{\hat{\Phi}} J(\hat{\Phi})\\[1ex]
\textnormal{subject to} \; \int_{\mathbb{R}^{n+q}_+} \hat{\Phi}(\xi) d\xi = 1  \\[1ex]
\hspace{20mm} \hat{\Phi}(\xi) \geq 0 \, \forall \xi\in\mathbb{R}^{n+q}_+,
\end{array}
\label{eq:op finite}
\end{equation}
where the last two constraints enforce that $\hat{\Phi}(\xi)$ is a probability distribution.

\begin{remark}
In the whole section the measured outputs $\Psi(\psi,t)$ are compared with the noise corrupted simulated output $\hat{\Psi}(\psi,t,\hat{\Phi})$. This is possible as we assume a large number of measured cells per measurement instant and therefore have good statistics on the measurement error.   
\end{remark}

\begin{figure}[t!]
\begin{center}
\includegraphics{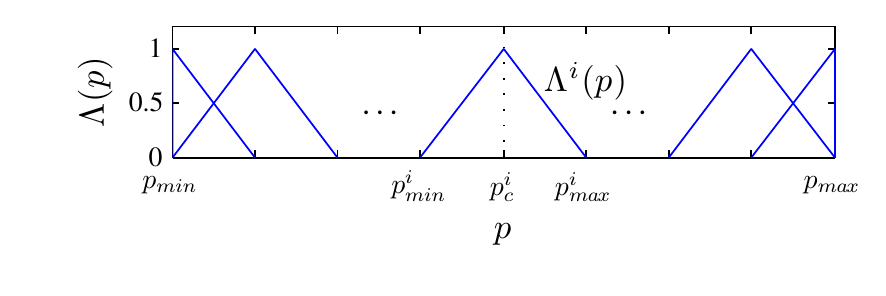}
\end{center}
\caption{Schematic of head functions $\Lambda^i(p)$.}
\label{fig:head functions}
\end{figure}

Unfortunately, the optimization problem \eqref{eq:op finite} is infinite dimensional. Therefore, a parametrization of $\hat{\Phi}$,
\begin{equation}
\begin{aligned}
\hat{\Phi}_\varphi(\xi) &= \sum_{i=1}^{n_{\varphi}} \varphi_i \Lambda^i(\xi),
\label{eq:parametrisation of Phi}
\end{aligned}
\end{equation}
with a weighting vector $\varphi \in \mathbb{R}^{n_\varphi}$ is introduced. In this work the ansatz functions $\Lambda^i$ for $\hat{\Phi}$ are chosen to be classical head functions, as depicted in Figure \ref{fig:head functions}. This yields the simplified, finite-dimensional optimization problem,
\begin{equation}
\begin{array}{l}
\varphi^* = \mathrm{arg} \min_{\varphi} J(\hat{\Phi}_\varphi)\\[1ex]
\textnormal{subject to } \; c^T \varphi = 1\\[1ex]
\hspace{24.5mm} \varphi \geq 0,
\end{array}
\label{eq:op finite 2}
\end{equation}
in which $c_i = \int_{\mathbb{R}^{n+q}} \Lambda^i(\xi) d\xi$. The two constraints are again needed to ensure that $\hat{\Phi}_\varphi(\xi)$ is a probability density.

In order to solve \eqref{eq:op finite 2} using computational techniques the quasi-linearity of \eqref{eq:PDE} is employed. As the superposition principle holds, the output $\hat{\Psi}(\psi,t,\hat{\Phi}_\varphi)$ can be written as the weighted sum
\begin{equation}
\begin{aligned}
\hat{\Psi}(\psi,t,\hat{\Phi}_\varphi) &= \sum_{i=1}^{n_{\varphi}} \varphi_i \hat{\Psi}(\psi,t,\Lambda^i),
\end{aligned}
\end{equation}
where $\hat{\Psi}(\psi,t,\Lambda^i)$ is the output distribution obtained for simulation with a parameter distribution according to $\Lambda^i(\xi)$. This allows the reformulation of the objective function to
\begin{equation}
\begin{aligned}
J\left(\hat{\Phi}_\varphi\right) = \sum_{k=1}^N \left|\left| \Psi(\psi,t_k) - \sum_{i=1}^{n_{\varphi}} \varphi_i \hat{\Psi}(\psi,t_k,\Lambda^i)\right|\right|_2^2.
\end{aligned}
\label{eq:l2 norm sum}
\end{equation}
Employing this the optimization problem \eqref{eq:op finite 2} can finally be written as
\begin{equation}
\begin{array}{l}
\varphi^* = \mathrm{arg} \min_{\varphi} \sum_{k=1}^N \left(A_k \varphi - b_k\right)^T W \left(A_k \varphi - b_k\right)\\[1ex]
\textnormal{subject to } \; c^T \varphi = 1\\[1ex]
\hspace{24.5mm} \varphi \geq 0,
\end{array}
\label{eq:op finite discretized}
\end{equation}
where the integral $||\cdot||_2^2$ is approximated, e.g. using the trapezoidal rule. The column vector $b_k$ contains hereby the values $\Psi(\psi,t_k)$ at the grid points of the discretization. Equivalently, the $i$th column of $A_k$ contains the values of $\hat{\Psi}(\psi,t_k,\Lambda^i)$ at the grid points. The matrix $W$ is a constant weighting matrix, determined by the chosen approximation of $||\cdot||_2^2$.

Note that problem \eqref{eq:op finite discretized} is convex. Hence, even in the case of high dimensional $\varphi$, convergence to the optimal parameter distribution within the considered class of distributions can be guaranteed.

\section{Application to the caspase cascade}
\label{sec:example}
Programmed cell death, also called apoptosis, is an important physiological process to remove infected, malfunctioning, or no longer needed cells from a multicellular organism. Pathways to induce apoptosis converge at the caspase activation cascade \citep{Hengartner2000}. A mathematical model for this network has been proposed by \cite{Eissing2004}. Here, we consider the caspase activation in response to an external death receptor stimulus, e.g. the tumor necrosis factor (TNF). As seen from experimental cytotoxicity assays, the cellular response to a TNF stimulus is highly heterogeneous, with some cells dying and others surviving. To understand the process at the physiological level it is thus crucial to consider the cellular heterogeneity, using for example cell population modeling.

\begin{figure}[t!]
\begin{center}
\includegraphics[width=10cm]{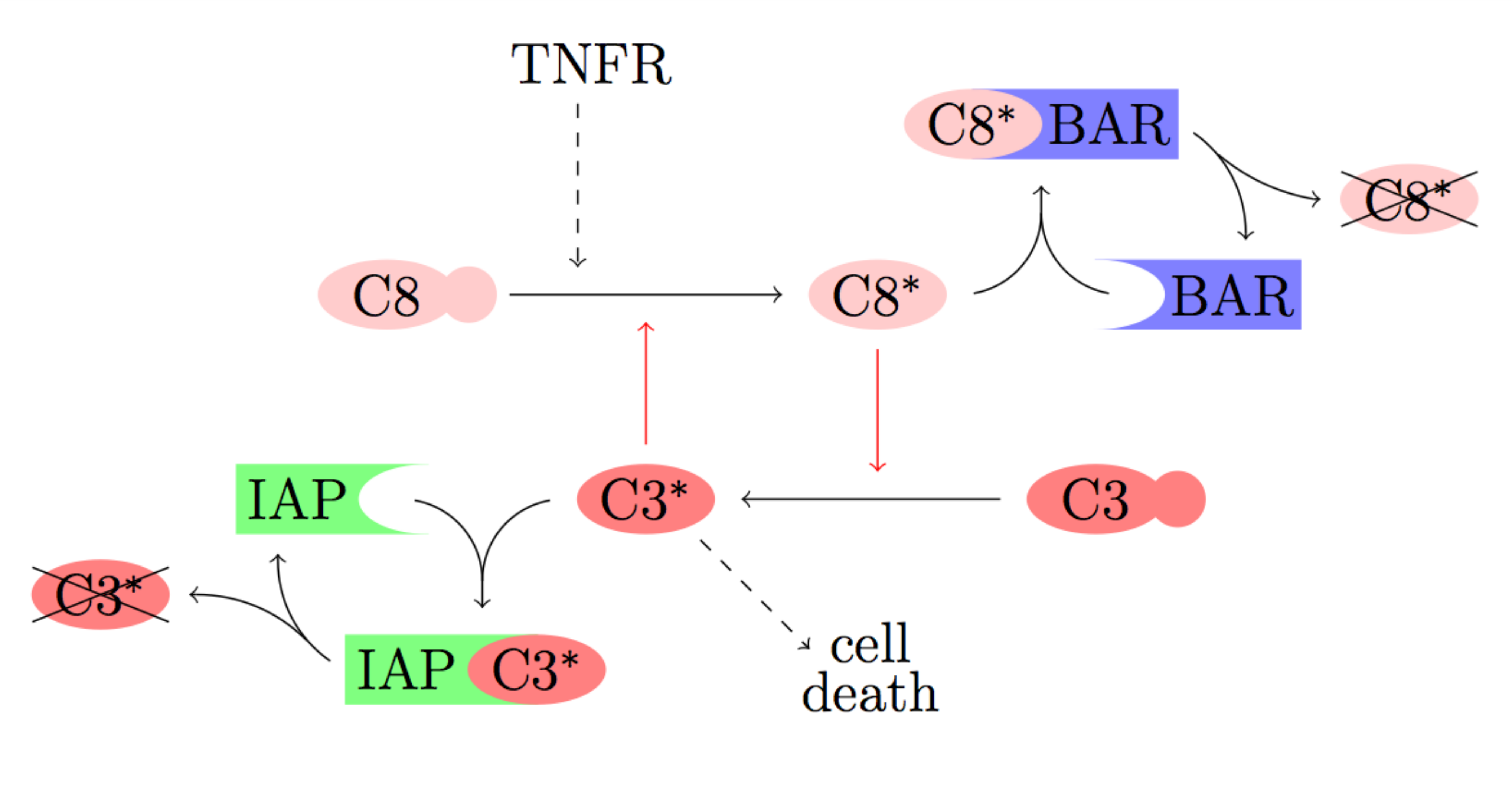}
\end{center}
\caption{Schematic of the caspase activation cascade.}
\label{fig:Caspase Network}
\end{figure}

The reactions for the single cell model are given by
\begin{eqnarray*}
  \textnormal{C3} + \textnormal{C8}^* &\overset{k_1}{\rightarrow}&  \textnormal{C3}^* + \textnormal{C8}^*  \\
  \textnormal{C3}^* + \textnormal{C8} &\overset{k_2}{\rightarrow}&   \textnormal{C3}^* + \textnormal{C8}^*  \\
  \textnormal{C3}^* + \textnormal{IAP} &\overset{k_3}{\underset{k_{-3}}{\leftrightarrows}}&   \textnormal{C3}^* \sim \textnormal{IAP}  \\
  \textnormal{C3}^* + \textnormal{IAP} &\overset{k_4}{\rightarrow}&   \textnormal{C3}^*  \\
  \textnormal{C8}^* &\overset{k_5}{\rightarrow}&   \emptyset  \\
  \textnormal{C3}^* &\overset{k_6}{\rightarrow}&   \emptyset  \\
  \textnormal{C3}^* \sim \textnormal{IAP} &\overset{k_7}{\rightarrow}&   \emptyset  \\
  \textnormal{IAP} &\overset{k_8}{\underset{k_{-8}}{\leftrightarrows}}&   \emptyset \\
  \textnormal{C8}  &\overset{k_9}{\underset{k_{-9}}{\leftrightarrows}}&   \emptyset \\
  \textnormal{C3}  &\overset{k_{10}}{\underset{k_{-10}}{\leftrightarrows}}&   \emptyset \\
  \textnormal{C8}^* + \textnormal{BAR} &\overset{k_{11}}{\underset{k_{-11}}{\leftrightarrows}}&   \textnormal{C8}^* \sim \textnormal{BAR} \\
  \textnormal{BAR}  &\overset{k_{12}}{\underset{k_{-12}}{\leftrightarrows}}&   \emptyset \\
  \textnormal{C8}^* \sim \textnormal{BAR} &\overset{k_{13}}{\rightarrow}&   \emptyset \\
  \textnormal{TNFR} + \textnormal{C8} &\overset{k_{14}}{\rightarrow}&   \textnormal{TNFR} + \textnormal{C8}^*
\end{eqnarray*}
For nominal parameter values, we refer to the original publication \citep{Eissing2004}. In comparison to the original model, we added reaction $v_{14}$ for the initiator caspase 8 (C8) activation by the TNF receptor complexes (TNFR). The reaction rate for this activation is given by $v_{14} = k_{14} [\textnormal{TNFR}] [C8]$, with the parameter value $k_{14} = 10^{-6} (\mathrm{molecules}\,\mathrm{min})^{-1}$. A sketch of the single cell model is given in Figure~\ref{fig:Caspase Network}.

\begin{figure}[t!]
\begin{center}
\includegraphics{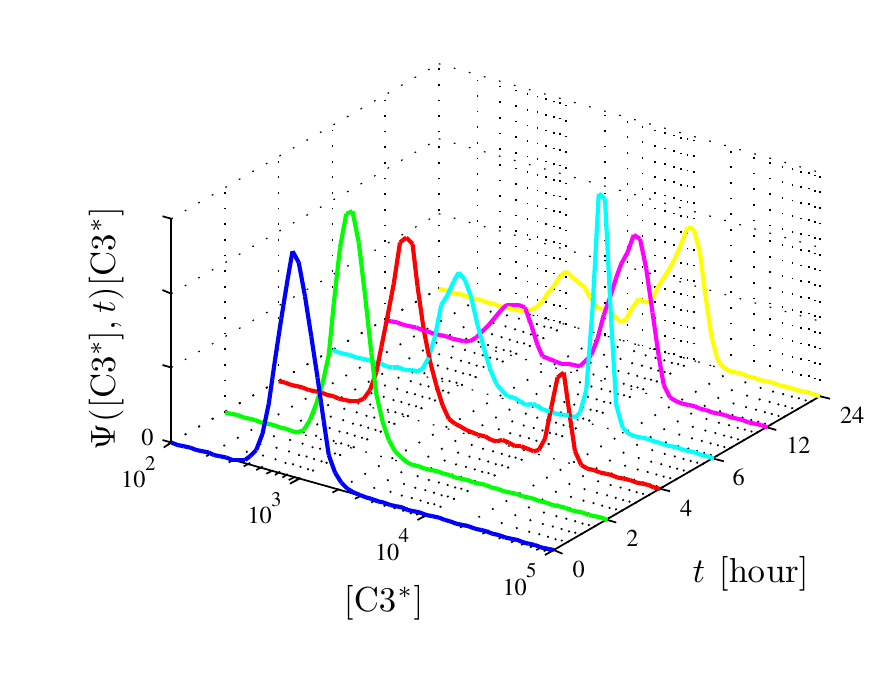};
\end{center}
\caption{Artificial noisy measurement data for amount of active caspase 3, [C3$^*$].}
\label{fig:ex meas}
\end{figure}

Heterogeneity is modeled by a $\mathrm{log}$-normally distributed production rate of the inhibitor of apoptosis IAP, $k_8$, and a $\mathrm{log}$-normally distributed amount of TNF-receptor complexes on the cell membrane, TNFR. These two quantities were chosen as it is known from experiments that there is a high cell-cell variability. Especially the concentration of IAPs contained in a cells is highly variable, and a variation in IAP production is known to affect cell death considerably \citep{EissingWal2006}. In the following the possibility of estimating the distributions of $\Phi(k_8)$ and $\Phi([\textnormal{TNFR}])$ from the distributions of [C3$^*$], $\Psi(\textnormal{C3}^*)$, is studied. The statistical model of the distribution, $\Psi(\textnormal{C3}^*)$ is shown in Figure~\ref{fig:ex meas}. This statistical model has been derived using artificial measurement data of $10^4$ cells at the measurement instances $t_k$, $k=1,\dots,6$. This is a realistic number for standard cytofluorometric experiments. The noise properties are assumed to be known and have been set to $\mu_1 = 0$, $\sigma_1 = 0.1$, $\mu_2 = \mathrm{log}(10^3)$, and $\sigma_2 = 0.3$, corresponding to an average measurement error of more than 20 percent.

\begin{figure}[t!]
\begin{center}
\includegraphics{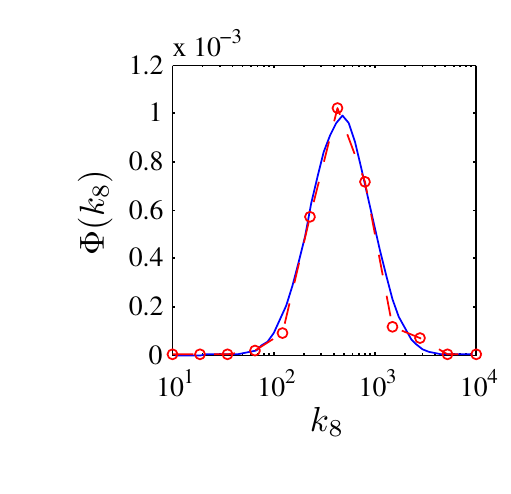}
\quad
\includegraphics{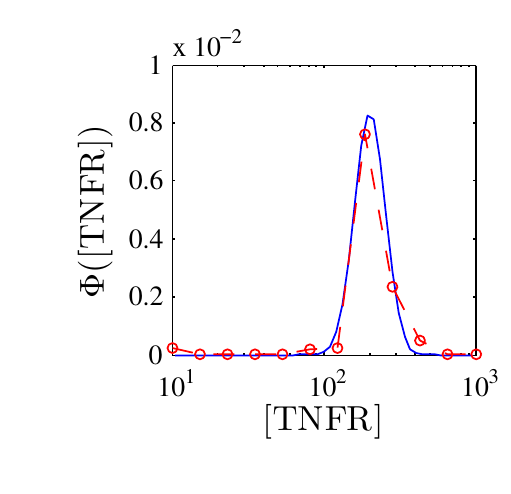}
\end{center}
\caption{Real ({\color{blue}---}) vs. estimated ({\color{red}-o-}) parameter distribution, with grid points ({\color{red}o}).}
\label{fig:ex est}
\end{figure}

Based on these data, the approach presented in Section~\ref{sec:estimation} is used to obtain an estimate for the parameter distribution. For this purpose the considered parameter set is divided using a 12 $\times$ 12 grid, with logarithmically distributed grid points. The grid points are used as edge and center points of the ansatz functions $\Lambda^i(k_8,[\textnormal{TNFR}])$ for $\hat{\Phi}(k_8,[\textnormal{TNFR}])$. The obtained estimation result is depicted in Figure~\ref{fig:ex est}.

It is obvious that the estimated parameter distribution approximates the real parameter distribution very well, especially considering the finite number of degrees of freedom. Hence, even though there is an average measurement error of 20 $\%$ on the single cell measurement, due to good statistics at the population level, the actual parameter distributions can be estimated accurately. Furthermore, this study shows that in principle, measuring one concentration can give enough information to estimate several parameter distributions, if the output distribution is sensitive with respect to these parameters.

\section{Summary and Conclusion}
\label{sec:conclusion}

Heterogeneity in cell populations is an important issue for research in systems biology. However, so far only few models describing heterogeneous populations of cells with more than one state variable have been developed. In this paper a partial differential equation model describing the time evolution of the state distribution is derived. We focused hereby in particular on the distribution of the measured outputs.

In the second part of the paper, the model of the noise corrupted measured outputs and its particular properties are used to estimate the parameter distributions underlying the heterogeneity. Therefore, a density-based statistical model of the sampled single cell used in combinations with $l_2$-norm based convex optimization.

Finally, we applied the developed estimation method to artificial data of a medium size bistable system modeling the caspase activation cascade. It could be shown that the proposed method yields good estimation results in case of a setup which is realistic in terms of noise and amount of available data.

\section*{Acknowledgments}

The authors acknowledge financial support from the German Federal Ministry of Education and Research (BMBF) within the FORSYS-Partner program (grant nr.\ 0315-280A), from the German Research Foundation within the Cluster of Excellence in Simulation Technology (EXC 310/1) at the University of Stuttgart, and from Center Systems Biology (CSB) at the University Stuttgart.



\begin{small}

\end{small}

\end{document}